\documentclass{ifacconf}

\usepackage{graphicx}      % include this line if your document contains figures
\usepackage{natbib}        % required for bibliography

\usepackage{amssymb}
\usepackage{epstopdf}
\usepackage{amsmath}
\usepackage{mathtools, cuted}
\usepackage{multirow}
\usepackage{pbox}
\usepackage{algorithm}
\usepackage{subfigure}
\usepackage{bm}
\usepackage{url}

\usepackage{booktabs}
\usepackage[table, dvipsnames]{xcolor}
\definecolor{VeryLightGray}{rgb}{0.93,.93,.93}

\usepackage{tabularx}
\usepackage{stackengine}
\setstackEOL{\\}

\newtheorem{problem}{\textbf{Problem}}
\newtheorem{definition}{\textbf{Definition}}

\newtheorem{lemma}{\textbf{Lemma}}
\newtheorem{remark}{\textbf{Remark}}

\newtheorem{proposition}{\textbf{Proposition}}

\newcommand*{\N}{\mathbb{N}}
\newcommand*{\R}{\mathbb{R}}

\begin{document}
\begin{frontmatter}

\title{Memoryless Cumulative Sign Detector for Stealthy CPS Sensor Attacks\thanksref{footnoteinfo}} 

\thanks[footnoteinfo]{This work is based on research sponsored by ONR under agreement number N000141712012, and NSF under grant \#1816591.}

\author{Paul J Bonczek and Nicola Bezzo} 

\address{Charles L. Brown Department of Electrical and Computer Engineering, and Link Lab, University of Virginia, Charlottesville, VA 22904, USA. (e-mail: \{pjb4xn, nb6be\}@virginia.edu).}

\begin{abstract}                % Abstract of not more than 250 words.
Stealthy false data injection attacks on cyber-physical systems introduce erroneous measurements onto sensors with the intent to degrade system performance. An intelligent attacker can design stealthy attacks with knowledge of the system model and noise characteristics to evade detection from state-of-the-art fault detectors by remaining within detection thresholds. However, during these hidden attacks, an attacker with the intention of hijacking a system will leave traces of non-random behavior that contradict with the expectation of the system model. Given these premises, in this paper we propose a run-time monitor called Cumulative Sign (CUSIGN) detector, for identifying stealthy falsified measurements by identifying if measurements are no longer behaving in a random manner. Specifically, our proposed CUSIGN monitor considers the changes in sign of the measurement residuals and their expected occurrence in order to detect if a sensor could be compromised.  Moreover, our detector is designed to be a memoryless procedure, eliminating the need to store large sequences of data for attack detection. We characterize the detection capabilities of the proposed CUSIGN technique following the well-known $\chi^2$ failure detection scheme. Additionally, we show the advantage of augmenting CUSIGN to the model-based Cumulative Sum (CUSUM) detector, which provides magnitude bounds on attacks, for enhanced detection of sensor spoofing attacks. Our approach is validated with simulations on an unmanned ground vehicle (UGV) during a navigation case study.
\end{abstract}


\end{frontmatter}
%===============================================================================

\section{Introduction} \label{sec:introduction}

Today's cyber-physical systems (CPSs) are fitted with multiple on-board sensors and computers that make them capable of many civilian and military applications with minimal/no human supervision. Autonomous navigation, transportation, surveillance, and task oriented jobs are becoming more common and ready for deployment in real world applications especially in the automotive, industrial, and military domains. These various enhancements in autonomy are possible thanks to the tight interaction between computation, sensing, communications, and actuation that characterize CPSs. With these increasing capabilities, comes the risk of more security vulnerabilities to cyber-attacks like sensor spoofing with the intent to induce undesired system behavior. An example of this problem was demonstrated by authors in [\cite{YachtSpoof}] in which GPS data were spoofed to slowly drive a yacht off the intended route. 

Many systems, including vehicle technologies, typically have well studied dynamical models and their sensors have specific expected behaviors according to their characterized noise profiles. Malicious attackers aim to compromise a system by diverting system states to unsafe regions, while remaining hidden within system detection boundaries. Despite lying within magnitude boundaries to remain undetected, non-random patterns arise that violate the expected behavior from normal system behavior. For example an attacker with the intention of hijacking an autonomous system while remaining stealthy will manipulate sensor measurements pushing them toward one direction.

Considering the problem at hand, in this work, we leverage the known characteristics of the \textit{residual} -- defined as the difference between sensor measurement and state prediction -- to build a memoryless run-time monitor to detect non-random behaviors in sensing. To this end, we consider the $\chi^2$ detection scheme [\cite{BadData}] which creates a {\em test measure} to monitor a vector of Normally distributed residuals. To monitor for randomness, we leverage the signed value of the difference between the $\chi^2$ distributed test measure and an arbitrary reference point within its known distribution. Systems operating under normal conditions have expected probabilities of whether the test measure should be greater or less than the chosen reference point. Our Cumulative Sign (CUSIGN) dynamic detector, inspired by Cumulative Sum (CUSUM) theory [\cite{CUSUM_Page}], leverages the history of sign valued differences between the test measure and the reference point, resulting in an alarm rate which is monitored at run-time for attack detection purposes. Thus, as a sensor is compromised, its corresponding residual will leave a trail of non-random behavior and will not follow an expectation.
 
In summary, the main objective of this work is to find stealthy sensor attacks exhibiting non-random behavior within the noise profile of a system in the presence of sensor and process noise. The contribution of this paper is twofold: 1) we propose the CUSIGN detection framework to deal with hidden non-random sensor attacks, typically undetectable by conventional detectors, by monitoring the expected alarm rate associated with consecutive changes of signs in the \textit{test measure}; 2) we introduce a memoryless feature to the CUSIGN detection procedure by leveraging a modified version of Welford's online algorithm [\cite{Welford}], which we call a {\em Memoryless Run-time Estimator} (MRE), that uses a pseudo-window to monitor the alarm rate at run-time, removing the need of storing the entire sequence of data over the duration of the operation. We show empirical results about the MRE with a chosen pseudo-window length to find bounds for detection. Our framework is also combined with the CUSUM technique to create a complete detector framework. Furthermore, we include simulations on a UGV model to validate the proposed detection scheme.

\subsection{Related Work}
\label{sec:Related Work}

The subject of CPS security has garnered considerable interest in analyzing detection methods for stealthy sensor attacks that intend to degrade system performance. This work builds on previous research considering deceptive cyber-attacks to systems by injecting false data to sensor measurements while trying to remain undetected [\cite{BadData}]. Previous works have analyzed the effects of malicious sensor attacks on the Kalman filter [\cite{KF_attack}]. Similarly, authors in [\cite{BadData,CST1}] discuss how undetected attacks can compromise closed-loop systems, causing state and system dynamic degradation.

Several attack detection techniques exist in the literature that analyze the residual, one of which is the Sequential Probability Ratio Testing (SPRT) [\cite{SPRT2}] that tests the sequence of incoming residuals one at a time by taking the log-likelihood function (LLF). Compound Scalar Testing (CST) in [\cite{CST1}] is a computationally friendly technique that reduces the residual vector with known residual variances into a scalar test measure of $\chi^2$ distribution. An improvement of CST in [\cite{CST2}] is made by including a coding matrix to sensor outputs that is unknown to attackers, then an iterative optimization algorithm is used to solve for a transform matrix to detect stealthy attacks.

Different from these previous works that leverage residual-based techniques, we build a framework to monitor sensor measurements to find previously undetectable attacks by searching for non-random behavior. The CUSIGN detector proposed in this work to find non-random behaviors is inspired by the theory of CUmulative SUM (CUSUM), developed in [\cite{CUSUM_Page}] that is commonly used as a monitor for change detection, such as a change in mean. Authors in [\cite{CUSUM_Journal}] formalized a model-based detector of the CUSUM algorithm by leveraging known characteristics of the system dynamical and noise models.

Several statistical techniques are available in literature to test for randomness by leveraging a sequence of data and test a hypothesis. Among randomness tests, the Wald-Wolfowitz runs test [\cite{wald1940}] observes consecutive values that belong to one of two different groups (known as a \textit{run}) over a given sequence. Similarly, the Serial Independence runs test [\cite{serial_test}] observes the number of runs of the difference between current and previous values over a sequence of data. The Monobit (frequency) test [\cite{Monobit}] observes a sequence of $1$'s and $0$'s to determine if they are equally probable. This work leverage these principles to detect non-random patterns in sensor measurements.

The remainder of this work is organized as follows: In Section \ref{sec:preliminaries} we begin by introducing the system modeling and problem formulation, followed by the characterization of our CUSIGN detector with an empirically derived memoryless detector operation to provide detection bounds in Section \ref{sec:CUSIGN}. In Section \ref{sec:CUSUM} we briefly discuss the CUSUM attack detector to compare with CUSIGN. Finally, in Section \ref{sec:Results} we demonstrate through simulations the performance of our framework augmented with an existing CUSUM detector before drawing conclusions in Section~\ref{sec:conclusion}.

\section{Preliminaries \& Problem Formulation} \label{sec:preliminaries}

\subsection{Model}
\label{sec:model}

In this work we consider autonomous cyber-physical systems whose dynamics are described by a discrete-time linear system in the following form:
\vspace{5pt}
\begin{equation}\label{eq:system1}
\begin{split}
\bm{x}_{k+1} &= \bm{A} \bm{x}_k + \bm{B} \bm{u}_k + \bm{\nu}_k \\
	\bm{y}_k&=\bm{C} \bm{x}_k + \bm{\eta}_k ,
\end{split}
\end{equation}

with $\bm{A} \in \R^{n\times n}$ the state matrix, $\bm{B} \in \R^{n\times m}$ the input matrix, and $\bm{C} \in \R^{s\times n}$ the output matrix with the state vector $\bm{x}_k \in \R^{n}$, system input $\bm{u}_k \in \R^{m}$, output vector $\bm{y}_k \in \R^{s}$ providing measurements from $s$ sensors from the set $\mathcal{\bm{S}}=\{1,2,\dots,s\}$, and sampling time-instants $k \in \N$. Process and measurement noises are multivariate zero-mean Gaussian uncertainties $\bm{\nu} = \mathcal{N}(0,\bm{Q}) \in \R^n$ and $\bm{\eta} = \mathcal{N}(0,\bm{R}) \in \R^s$ with covariance matrices $\bm{Q} \in \R^{n\times n}, \bm{Q} \geq 0$ and $\bm{R} \in \R^{s\times s}, \bm{R} \geq 0$ respectively, and are assumed static.

During operations, a Kalman Filter (KF) is implemented to provide a state estimate $\hat{\bm{x}}_k \in \R^n$ in the following form,
\vspace{-1.5pt}
\begin{equation}\label{eq:Kalman}
	\hat{\bm{x}}_{k+1} = \bm{A} \hat{\bm{x}}_k + \bm{B} \bm{u}_k + \bm{L}(\bm{y}_k - \bm{C}\hat{\bm{x}}_k), 
\end{equation}

where the Kalman gain matrix $\bm{L} \in \R^{n \times s}$ is 
\vspace{3.5pt}
\begin{equation}
\label{eq:SteadyState_P_K}
	\bm{L} = \bm{A}\bm{P}\bm{C}^T(\bm{C}\bm{P}\bm{C}^T + \bm{R})^{-1}.
\end{equation}

For ease, we assume that the KF is at steady state before sensor attacks occur, such that $\lim_{k\to \infty} \bm{P}_k~=~\bm{P}$. The estimation error of the steady state KF is defined as $\bm{e}_k = \bm{x}_k - \hat{\bm{x}}_k$ while its {\em residual} $\bm{r}_k$ is given by
\vspace{3.5pt}
\begin{equation}
\label{eq:Residual}
	\bm{r}_k = \bm{y}_k - \bm{C}\hat{\bm{x}}_k = \bm{C}\bm{e}_k + \bm{\eta}_k + \bm{\xi}_k,
\end{equation}

and the covariance matrix of the residual \eqref{eq:Residual} is defined as
\vspace{-.5pt}
\begin{equation}
\label{eq:Residual_Covariance}
	\bm{\Sigma} = \mathrm{E}[\bm{r}_{k+1}\bm{r}_{k+1}^T]  =  \bm{C}\bm{P}\bm{C}^T + \bm{R} \in \R^{s \times s}.
\end{equation}

\subsection{Residual for Detection} 
\label{sec:residue}

A widely used sensor measurement failure detector in CPSs is the $\chi^2$ detector [\cite{BadData}], computed by the following quadratic test measure
\vspace{3pt}
\begin{equation}
\label{eq:z_scalar}
    z_k = \bm{r}_k^T\bm{\Sigma}^{-1}\bm{r}_k = \bm{r}_k^T(\bm{CPC}^T+\bm{R})^{-1}\bm{r}_k \in \R^{\geq 0}.
\end{equation}

In the absence of sensor attacks, the residual is a Normally distributed random vector $\bm{r}_k \sim \mathcal{N}(0,\bm{\Sigma})$ where $\bm{r}_k \in \R^s$, the test measure $z_k$ belongs to a $\chi^2$ distribution with $s$ degrees of freedom. Under the assumption that the system is not under attack (i.e. the residual satisfies \eqref{eq:Residual_Covariance}), the scalar test measure in \eqref{eq:z_scalar} follows
\vspace{3pt}
\begin{equation}
    \label{eq:Expected_z}
	\mathrm{E}[z_{k}] = s, \hspace{12pt} \mathrm{Var}[z_{k}] = 2s.
\end{equation} 

The general case of the $\chi^2$ detector compares the scalar test measure $z_k$ to a threshold $T$ by:
\vspace{2pt}
\begin{equation}
\begin{split}
    \label{eq:Chi_square_thresh}
    \bigg\{ \begin{array}{lll}
	z_{k} \leq T & \longrightarrow & \textit{no alarm}, \\
    z_{k} > T & \longrightarrow & \textit{alarm}.
    \end{array}
\end{split}
\vspace{-3pt}
\end{equation} 

where the design of the threshold $T$ is independent of system noises and based on the number of sensors, $s$.

When one or more sensors are attacked, properties of the residual $\bm{r}_k$ and the test measure $z_k$ may no longer hold. When considering sensor attacks, the output of the system can be written as:
\vspace{3pt}
\begin{equation}\label{eq:output_equation}
	\bm{y}_k = \bm{C} \bm{x}_k + \bm{\eta}_k + \bm{\xi}_k ,
\end{equation}

where $\bm{\xi}_k \in \R^s$ represents the attack vector subject to false data injection attacks onto sensors. A limitation to the method from \eqref{eq:Chi_square_thresh} is that stealthy attacks purposely hidden within detection boundaries may be undetectable. However, an attacker with the intent of hijacking a CPS, may leave traces of non-random behavior on the test measure $z_k$. To detect such non-random behavior, we propose a framework consisting of adding a memoryless run-time dynamic detector on the test measure $z_k$ searching for non-random behavior, while eliminating the need to store large amounts of data for detection purposes.

\subsection{Problem Statement}
\label{sec:problem}
An attacker trying to hijack a system, will consequently leave behind non-random behavior to sensor measurements. In this work we focus specifically on sign changes and their expected occurrences. With these considerations in mind, a system that is not compromised will have measurement residuals with signs that are normally distributed and with proper rate of sign changes. 

\begin{definition}\label{random_definition}
A system that is not compromised will behave in a random manner if the signed value of the difference between the test measure and a reference point maintain an expected sign occurrence.
\end{definition}
Since we are considering sensor spoofing, unknown attack signals containing malicious data can disrupt randomness, resulting in measurements that display non-random signed behavior. Formally, the problem that we are interested in solving is:
\begin{problem} 
\label{problem1} {\textbf{Randomness of Measurements:}} 
Given the test measure \eqref{eq:z_scalar} computed from the residual $\bm{r}_k$ and the residual covariance $\bm{\Sigma}$ as defined in \eqref{eq:Residual} and \eqref{eq:Residual_Covariance}, find a policy to determine at run-time whether a sensor measurement is non-random, i.e., if the condition in Definition \ref{random_definition} does not hold.
\end{problem}

Furthermore, in this work we impose that the computation from all aspects of the detector must have a memoryless property.
\begin{definition}\label{memoryless_definition}
A detector satisfies the memoryless property when the detection procedure does not rely on storing and using a sequence of data over any window horizon.
\end{definition}

\begin{problem} 
\label{problem2} {\textbf{Memoryless Property:}} 
With the given test measure \eqref{eq:z_scalar} to analyze, find a policy for a memoryless detection procedure without the need to store a collection of data over any length of time to determine if the system is compromised, satisfying the condition in Definition \ref{memoryless_definition}.
\end{problem}

\vspace{-3pt}
\begin{figure}[th!b]
\centering
\includegraphics[width=0.48\textwidth]{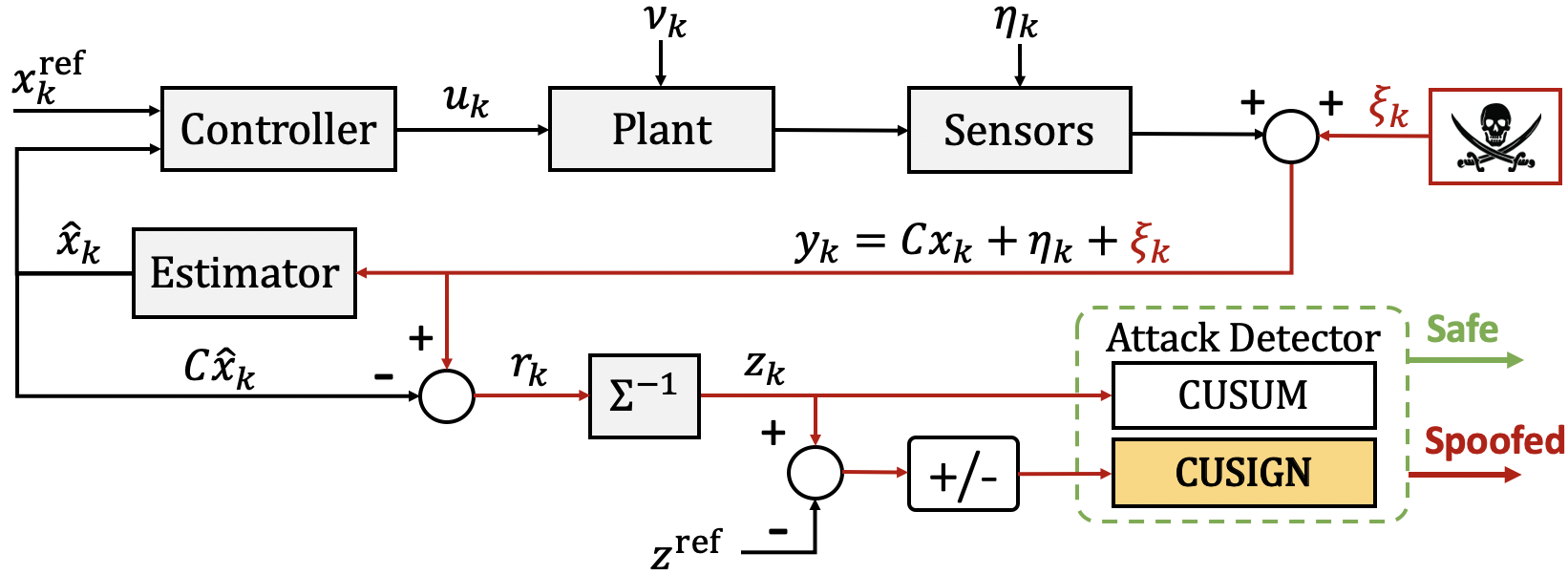}
\vspace{-6pt}
\caption{The architecture of a CPS while experiencing sensor attacks augmented with our CUSIGN detector.}
\label{fig:AttackDiagram}
\vspace{2pt}
\end{figure}

\subsection{System Architecture}
\label{sec:system_arch}

The overall cyber-physical system architecture including the CUSIGN detector is summarized in Fig.~\ref{fig:AttackDiagram}. CUSIGN, which can be augmented to any boundary detector providing magnitude bounds, is placed in the system feedback to monitor the relationship between measurement and state prediction. We focus on stealthy sensor attacks where an attacker may inject an attack signal at any point between the sensors and the state estimator, in an attempt to affect system behavior. 

\section{Cumulative Sign Detector} \label{sec:CUSIGN}

We develop a Cumulative Sign (CUSIGN) detector that analyzes the sign of the given test measure $z_k$ relative to a reference point and determines whether there is non-random behavior occurring. The model-based CUSIGN detector monitors the test measure from \eqref{eq:z_scalar} and outputs an alarm when the CUSIGN test variable reaches a user defined threshold. For a given user defined threshold, an expected alarm rate can be found that is independent from the model of the system \eqref{eq:system1}.
% \NB{next sentence needs still improvements} 

In normal conditions, i.e., without attacks or sensor malfunctions, the test measure $z_k$ has a specific probability of being higher or lower than a given user defined reference point $z^{\textrm{ref}} \in \R^{> 0}$ within its known distribution. We formalize these probabilities of $z_k$ being higher or lower than the reference point by
\begin{equation}
\label{eq:prob_z_mean}
    \begin{split}
    & \mathrm{Pr}\big(z_k < z^{\textrm{ref}}\big) = \gamma\Big(\frac{s}{2}, \frac{z^{\textrm{ref}}}{2}\Big), \\
    & \mathrm{Pr}\big(z_k > z^{\textrm{ref}}\big) = 1- \gamma\Big(\frac{s}{2}, \frac{z^{\textrm{ref}}}{2} \Big),
    \end{split}
\end{equation}

where $\gamma(\cdot,\cdot)$ is the \textit{regularized lower incomplete gamma function} [\cite{statsbook}]. The sign of $z_k$ with respect to the reference $z^{\textrm{ref}}$ is computed by the following
\vspace{1pt}
\begin{equation}
\begin{split}
    \label{eq:Sign_function}
    \mathrm{sgn}(z_k-z^{\textrm{ref}}) :=\Bigg\{ \begin{array}{ll}
	-1, & \hspace*{2pt} \textbf{if } z_{k}-z^{\textrm{ref}} < 0, \\
    0, & \hspace*{2pt} \textbf{if } z_{k}-z^{\textrm{ref}} = 0, \\
    1, & \hspace*{2pt} \textbf{if } z_{k}-z^{\textrm{ref}} > 0, \\
    \end{array}
\end{split}
\end{equation} 
\vspace{-6pt}

where the probability of each scenario occurring is
\vspace{1pt}
\begin{equation}
\label{eq:binomial_probs}
    \begin{split}
    \mathrm{Pr}\big(\text{sgn}(z_k-z^{\textrm{ref}}) = -1 \big) &= p_-, \\
    \mathrm{Pr}\big(\text{sgn}(z_k-z^{\textrm{ref}}) = 0 \big) &= 0, \\
    \mathrm{Pr}\big(\text{sgn}(z_k-z^{\textrm{ref}}) = 1 \big) &= p_+.
    \end{split}
\end{equation}
\vspace{-5pt}

\begin{figure}[th!b]
\centering
\includegraphics[width=0.47\textwidth]{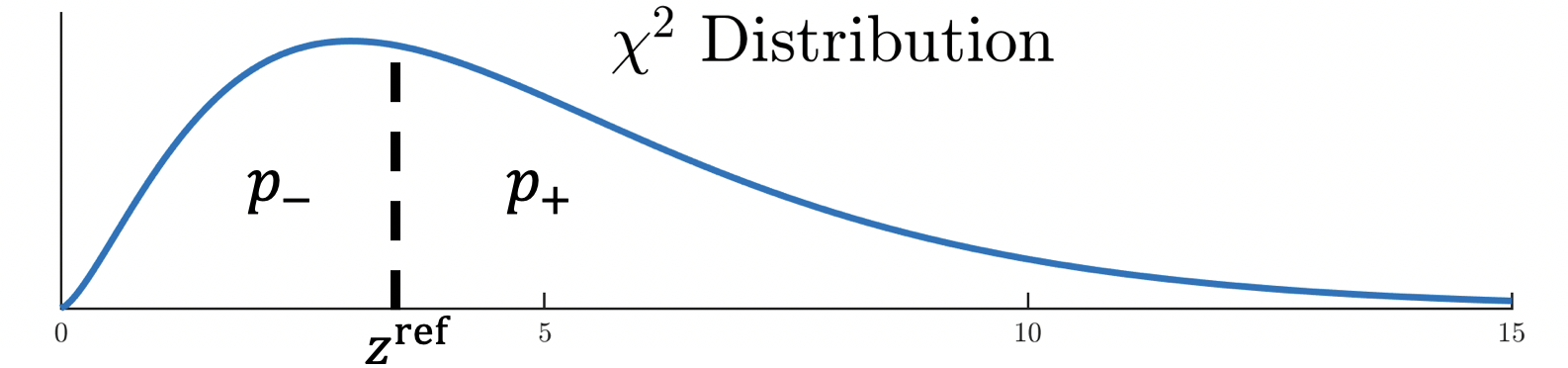}
\vspace{-4pt}
\caption{Probabilities $p_+$ and $p_-$ determined by $z^{\textrm{ref}}$.}
\label{fig:chi2}
\vspace{-1pt}
\end{figure}

An example of \eqref{eq:binomial_probs} is shown in Fig. \ref{fig:chi2}, where the probabilities $p_+$ and $p_-$ determine whether $z_k$ will be higher or lower than $z^{\textrm{ref}}$ given $z_k \sim \chi^2$.

The procedure of CUSIGN is an accumulation of signed values, denoted by the CUSIGN test variables $S_k^+$ and $S_k^-$. Each variable is a monitor checking for a change in the probability of the signed value $ \mathrm{sgn}(z_k-z^{\textrm{ref}})$, one for {\em positive} and the other for {\em negative} changes. The following procedure summarizes the CUSIGN detection in both the positive and negative cases:
\vspace{9pt}

\centerline{\textbf{CUSIGN Detector Procedure}}
\vspace{2pt}
\hrule
\vspace{-6pt}
  \begin{equation}
  \label{pro:CUSIGN_pos}
      \begin{array}{ll}
        \hspace*{-7pt} \textbf{Initialize } S_{0}^+ = 0,  & \\
        \hspace*{-7pt} S_{k}^+ = \max \hspace*{-2pt} \big( 0,S_{k-1}^+ \hspace*{-1pt} + \text{sgn}(z_k-z^{\textrm{ref}}) \big) , &  \\
        \vspace{4pt}
        \hspace*{-7pt} S_{k}^+ = 0 \text{ and Alarm } \zeta_k^+ = 1, & \textbf{if } S_{k-1}^+ \hspace*{-1pt} = \tau, \\
        \hspace*{-7pt} \textbf{Initialize } S_{0}^- = 0, & \\
        \hspace*{-7pt} S_{k}^- = \min \hspace*{-2pt}  \big( 0,S_{k-1}^- + \text{sgn}(z_k-z^{\textrm{ref}}) \big) , &  \\
        \hspace*{-7pt} S_{k}^- = 0 \text{ and Alarm } \zeta_k^- = 1, & \textbf{if } S_{k-1}^- \hspace*{-1pt} = \hspace*{-1pt} -\tau.
      \end{array}
  \end{equation}
  \vspace{-1pt}
  \hrule

\begin{figure}[b!ht]
\centering
\includegraphics[width=.96\linewidth]{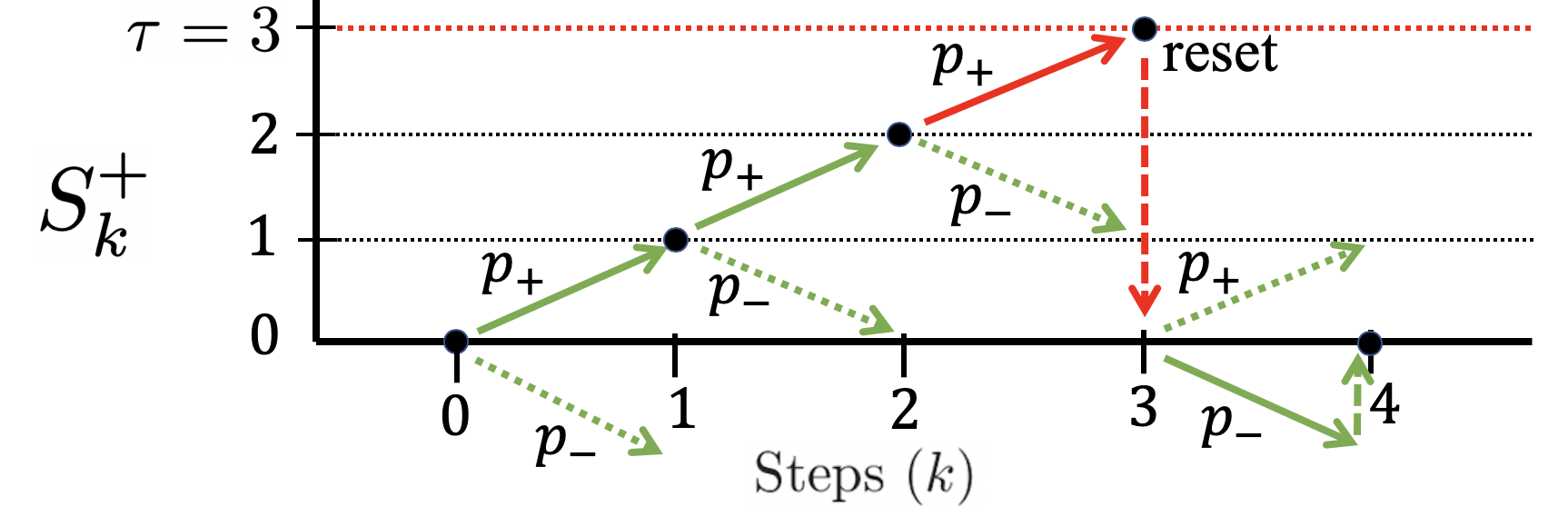}
\vspace{-6pt}
\caption{Transitions of the CUSIGN test variable $S_k^+$ with threshold $\tau = 3$.}
\label{fig:CUSIGN_trans}
\end{figure}

The design of the test variable sequences $S_k^+$ and $S_k^-$ are to accumulate the signed value $\text{sgn}(z_k-z^{\textrm{ref}}) \in \{-1,0,1 \}$ and triggering an alarm $\zeta_k^{\pm} = \{ \zeta_k^{+},\zeta_k^{-} \} \in \{0,1 \}$ when the test variables reach the threshold values $ \tau \in \N^{+}$. When either of the test variables are equal to their corresponding thresholds, the given test variable is reset to $0$. An example of the CUSIGN test variable is shown in Fig. \ref{fig:CUSIGN_trans} where three consecutive iterations $z_k > z^{\textrm{ref}}$ are satisfied at $k = 1,2,3$ (transitioning $S_k^+$ in the direction of $p_+$). At $k=3$, the CUSIGN test variable $S_k^+$ reaches the threshold value $\tau=3$ causing a reset such that $S_k^+ \rightarrow 0$.

Choosing a specific threshold $\tau$ results in expected alarm rates $\mathrm{E}[\alpha^{+}]$ and $\mathrm{E}[\alpha^{-}]$ for both the positive and negative cases of the CUSIGN procedure \eqref{pro:CUSIGN_pos}. In the case that $z^{\textrm{ref}} = \mathrm{E}[\text{median}(z_k)]$ such that $p_+ = p_-$, the resulting expected alarm rates are equal $\mathrm{E}[\alpha^{+}] = \mathrm{E}[\alpha^{-}]$.

\begin{figure}[b!ht]
\centering
\includegraphics[width=.93\linewidth]{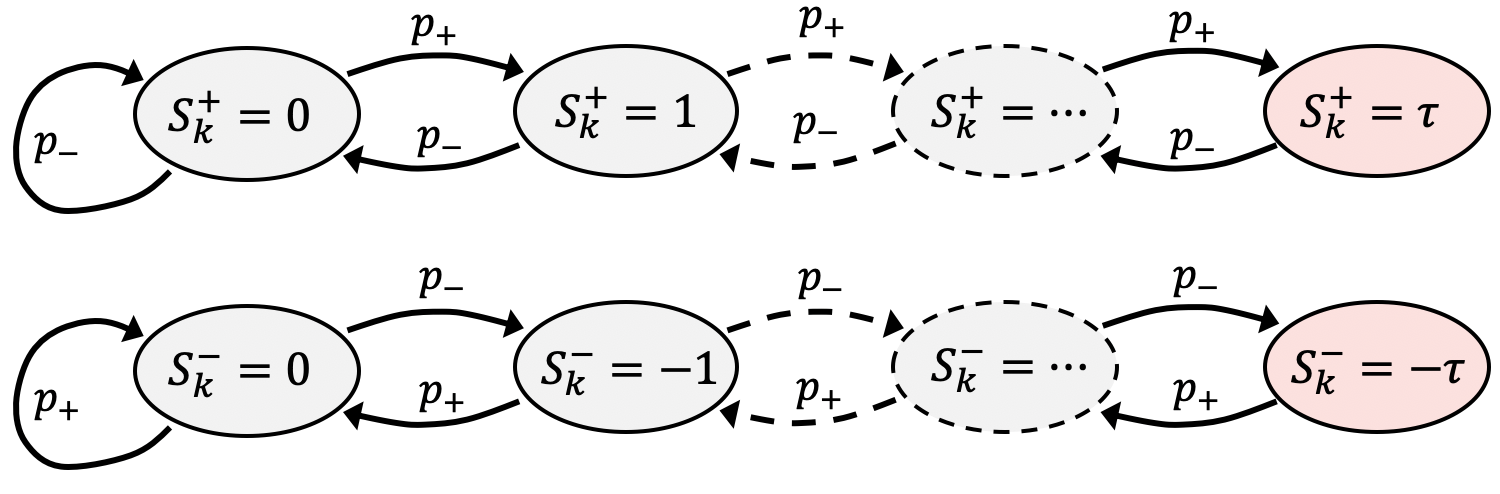}
\vspace{-9pt}
\caption{Markov chain for both positive and negative cases of the CUSIGN test sequence with threshold $\tau$.}
\label{fig:Markov_trans}
\vspace{-2pt}
\end{figure}

Similar to the implementation in [\cite{CUSUM_Journal}], the transition of the CUSIGN test sequences $S_k^{\pm}$ can be constructed as a Markov chain with a transition matrix modeled from the probabilities of $\text{sgn}(z_k - z^{\textrm{ref}})$. With a user defined threshold $\tau$ to trigger an alarm and causing a reset condition of the CUSIGN test variable to $0$, we show the transitions of $S_k^{\pm}$ with a Markov chain diagram, as follows in Fig. \ref{fig:Markov_trans}.
\vspace{-1pt}

Given a chosen threshold value $\tau \in \N^+$ as a value that triggers an alarm when $|S_k^{\pm}| = \tau$, we describe the Markov chain in Fig. \ref{fig:Markov_trans} in the form of a Markov transition matrix $\mathcal{T^{\pm}} \in \R^{(\tau+1) \times (\tau+1)}$. The CUSIGN Markov Chain, occurring in a discrete manner, contains $\tau+1$ states denoted as $\mathcal{M} = \{M_0,M_1,\dots,M_{\tau} \}$ where $M_{\tau}$ is an absorbing state that is equal to the threshold, causing the CUSIGN test sequence $S_k^{\pm}$ to reset to $M_0$ (i.e., $S_k^{\pm}~=~0$). The CUSIGN Markov transition matrix $\mathcal{T}^{\pm}$ for both positive $\mathcal{T}^{+}$ and negative $\mathcal{T}^{-}$ cases with a probability distribution of $\text{sgn}(z_k - z^{\textrm{ref}})$ are written by
\vspace{2pt}
\begin{equation}
\setlength\arraycolsep{5pt}
\label{eq:T_transition}
    \mathcal{T}^{\pm} = \begin{bmatrix} p_{\mp} & p_{\pm} & 0 & 0 & \dots & 0 \\ p_{\mp} & 0 & p_{\pm} & 0 & \dots & 0 \\ 0 & p_{\mp} & 0 & p_{\pm} &  & 0 \\ \vdots &  & \ddots & & \ddots & \vdots \\ 0 & \dots & 0 & p_{\mp} & 0 & p_{\pm} \\ 0 & \dots & 0 & 0 & 0 & 1 \end{bmatrix}.
\end{equation}
\vspace{-4pt}

The transition matrix $\mathcal{T^{\pm}}$ structure remains the same on any system, where the matrix size depends only on the value of the threshold $\tau$. Transition probabilities for transient states in $\mathcal{T}^{\pm}$ adhere to the following
\vspace{3pt}
\begin{equation}
\label{eq:transprob_CUSIGN}
    \hspace*{-5pt} \mathcal{T}^{\pm} \hspace*{3pt} \Bigg\{ \begin{array}{ll}
    \mathrm{Pr}(M_j \rightarrow M_{j+1}) = p_{\pm}, & \text{for } j \hspace*{-1pt} = \hspace*{-1pt} \{0,\dots,\tau \hspace*{-1pt} -\hspace*{-1pt} 1 \}, \\
    \mathrm{Pr}(M_j \rightarrow M_{j-1}) = p_{\mp}, & \text{for } j \hspace*{-1pt} = \hspace*{-1pt} \{1,\dots,\tau \hspace*{-1pt} - \hspace*{-1pt} 1 \}, \\
    \mathrm{Pr}(M_0 \rightarrow M_0) = p_{\mp}, &  \\
    \end{array}
\end{equation}
\vspace{-4pt}

and the final row represents an absorbing state containing elements equal to $0$ besides the last element equaling $1$.

We define $\mathcal{R^{\pm}} \in \R^{\tau \times \tau}$ as a matrix obtained from $\mathcal{T^{\pm}}$ with its last row and column removed (i.e., the absorbing state at threshold $\tau$ is removed), representing the transition probabilities to and from the transient states, also known as the fundamental matrix. Elements of $\mathcal{R^{\pm}}$ are all non-negative and row sums are equal to or less than one, while the eigenvalues satisfy $\rho[\mathcal{R^{\pm}}] < 1$ such that $\mathcal{(R^{\pm}})^i \to 0$ as $i \to \infty$ and $\sum_{i=0}^{\infty} \mathcal{(R^{\pm}})^i = (\bm{I}_{\tau}-\mathcal{R^{\pm}})^{-1}$, where $\rho[\cdot]$ is the spectral radius and $\bm{I}_{\tau}$ is the identity matrix of size $\tau$. 

\begin{lemma}

Given a system with a CUSIGN detector \eqref{pro:CUSIGN_pos} with a chosen threshold $\tau \in \N^+$ and reference point $z^{\textrm{ref}} \in \R^{>0}$ that is not affected by sensor attacks such that the residual sequence satisfies $\bm{r}_k \sim \mathcal{N}(0, \bm{\Sigma}) \in \R^s$ and $z_k = \bm{r}_k^T \bm{\Sigma}^{-1} \bm{r}_k \sim \chi^2$ with $s$ degrees of freedom, then the inverse of the first element of the following vector
\vspace{2pt}
\begin{equation}
\label{eq:mu}
    \bm{\mu}^{\pm} = (\bm{I}_{\tau} - \mathcal{R^{\pm}})^{-1}\bm{1}_{\tau \times 1} = (\mu_1^{\pm}, \dots, \mu_{\tau}^{\pm})^T,
    \vspace{-3pt}
\end{equation}

is the expected alarm rate, i.e., $\mathrm{E}[\alpha^{\pm}] = (\mu_1^{\pm})^{-1}$.

\end{lemma}

\begin{pf}
    Given the Markov chain containing $\tau + 1$ states denoted by $\mathcal{M} = \{ M_0,M_1,\dots , M_{\tau} \}$, a fundamental matrix $\mathcal{R}^{\pm}$ is taken from a designed Markov transition matrix \eqref{eq:T_transition} to satisfy the transition probabilities \eqref{eq:transprob_CUSIGN}. Leveraging the theory of average run length (ARL) in CUSUM [\cite{CUSUM_ARL}], the ARL is defined as the average length of time for the test sequence to reach the threshold $\tau$ to trigger an alarm, determined by the fundamental matrix $\mathcal{R}^{\pm}$ containing the transient states within $\mathcal{T}^{\pm}$. By definition, the inverse of the ARL to observe an alarm results in the average frequency of obtaining an alarm, known as the alarm rate. The ARL can be found by computing \eqref{eq:mu}, then by inverting the first element of $\bm{\mu}^{\pm}$, i.e., $(\mu^{\pm}_1)^{-1}$, we obtain the expected alarm rate $\mathrm{E}[\alpha^{\pm}]$. 
\end{pf}

\subsection{Memoryless Run-time Estimation of Alarm Rates} \label{sec:AR_memoryless}

In the design of CUSIGN, we trigger an alarm when a test variable reaches a chosen threshold $\tau$. Given a system not experiencing sensor attacks, we have an expectation of the alarm rates. Typically, to find an alarm rate, the number of triggered alarms are tallied over a given period of time. In this work, we want to create a ``memoryless" procedure to find an alarm rate. 

The conventional method of finding an average $\bar{x}$ of a stochastic variable is $\bar{x}_{n} = \frac{1}{n} \left[ \sum_{i=1}^n x_i \right] $
where $n$ is the size the data set. This procedure requires storage of the complete data set, where computation becomes less efficient as $n$ grows. A memoryless online algorithm known as Welford's online algorithm for computing a mean incrementally was developed in [\cite{Welford}] by transforming the conventional method into an online update by the following form
\vspace{2pt}
\begin{equation}
\label{eq:Welford1}
    \begin{split}
        \hspace*{-5pt} \bar{x}_{n} & = \frac{1}{n} \left[x_n + \sum_{i=1}^{n-1} x_i \right] = \frac{1}{n} \left[ x_n + (n-1)\bar{x}_{n-1} \right] \\
        & = \frac{1}{n} \left[ x_n + n\bar{x}_{n-1} \hspace*{-1pt} - \hspace*{-1pt} \bar{x}_{n-1} \right] \hspace*{-1pt} = \bar{x}_{n-1} \hspace*{-1pt} + \hspace*{-1pt} \frac{\big[x_n \hspace*{-1pt} - \bar{x}_{n-1} \big]}{n}.
    \end{split}
\end{equation}

It can be seen in \eqref{eq:Welford1} that $n$ grows indefinitely, equal to the number of data points. We set a maximum value for $n$ such that $\max(n) = \ell \in \N^+$ to create a ``pseudo-window" for a rolling sequential estimation of an expected mean. We name this modified version of Welford's online algorithm utilizing a pseudo-window $\ell$ as a Memoryless Run-time Estimator (MRE). The behavior of MRE when computing the mean similarly imitates the conventional method of calculating the mean consisting of $\ell$ data points, but without the need to store the entire sequence.

For the case of attack detection using alarm rates for CUSIGN, we leverage MRE in \eqref{eq:Welford1} to find an online estimation of an expected alarm rate $\mathrm{E}[\alpha]$ (we omit $\pm$ for $\alpha^{\pm}$ in this section as the MRE applies to both the positive and negative cases). Leveraging the pseudo-window of length $\ell$ and replacing the counter $n$ from \eqref{eq:Welford1} with $k$ for sampling time instances, we attain the equation
\vspace{5pt}
\begin{equation}
\label{eq:alpha_est1}
    \hat{\alpha}_{k} = \hat{\alpha}_{k-1} + \frac{\big[\zeta_k - \hat{\alpha}_{k-1} \big]}{\ell},
\end{equation}

where $\zeta_k$ is the triggered alarm for CUSIGN, $\hat{\alpha}_k$ is an estimate of the alarm rate at time instance $k$, and $\hat{\alpha}_0 = 0$ initially at $k=0$.

\begin{proposition}
\label{prop:1}
Assuming the system is not experiencing sensor attacks and the test measure follows $z_k \sim \chi^2$ for time instances $k \geq 0$, we empirically find that the alarm rate is a Normal distribution as follows
\vspace{7pt}
\begin{equation}
\label{eq:AR_distribution}
    \hat{\alpha} \sim \mathcal{N}\bigg(\mathrm{E} [\alpha],\frac{\theta \mathrm{E} [\alpha](1-\mathrm{E}[\alpha])}{\ell}\bigg),
\end{equation}

where $\ell$ is the user defined pseudo-window length, $\theta \in \R^{>0}$ is an empirically found scaling value, and $\mathrm{E}[\alpha]$ is the expected alarm rate, i.e., the probability that the test variable $S_k$ reach the threshold, triggering an alarm $\zeta_k = 1$. 
\end{proposition}

Given the distribution of $\hat{\alpha}$ in Proposition \ref{prop:1}, the expectation of the estimated alarm rate follows
\vspace{5pt}
\begin{equation}
\label{eq:Expected_alpha}
    \mathrm{E} [\hat{\alpha}] = \mathrm{E} [\alpha], \hspace*{10pt} \mathrm{Var}[\hat{\alpha}] = \frac{\theta \mathrm{E} [\alpha](1 \hspace*{-1pt} - \hspace*{-1pt} \mathrm{E} [\alpha])}{\ell}.
    \vspace{2pt}
\end{equation}

Values of $\theta$ are found to be dependent on the chosen threshold $\tau$. Observed approximates of $\theta$ are presented in Table \ref{table:theta} for thresholds $\tau = 1,2,3,4$ and $\ell \geq 10$.

\begin{table}[htb!]
\caption{Empirical values for the scaling value $\theta$ given thresholds $\tau = 1,2,3,4$.}
\vspace{-5pt}
\centering
\begin{tabular}{ |p{1.5cm}||p{1.1cm}|p{1.1cm}|p{1.1cm}|p{1.1cm}|  }
 \hline
 \rule{0pt}{1.2\normalbaselineskip} Thresholds  & \centering $\tau = 1$ & \centering $\tau = 2$ & \centering $\tau = 3$ & \hspace*{3pt} $\tau = 4$ \\[2pt]
 \hline
 \rule{0pt}{1\normalbaselineskip} \centering $\theta$   & \centering $\frac{\ell}{2\ell-1}$ & \centering $\frac{.74\ell}{2\ell-1}$ & \centering $\frac{.7\ell}{2\ell-1}$ & \hspace*{5pt} $\frac{.69\ell}{2\ell-1}$ \\[1.7pt]
 \hline
\end{tabular}
\label{table:theta}
\end{table}
\vspace{5.5pt}

\begin{remark}
\label{rem:remark1}
    For the CUSIGN detector, we empirically find that $\hat{\alpha}$ follows \eqref{eq:AR_distribution} when $p_+ \approx p_-$ (i.e., $z^{\textrm{ref}}$ is chosen to be at $\mathrm{E} [\text{median}(z_k)]$ such that $p_- = p_+ = 0.5$). For a reference point $z^{\textrm{ref}}$ not placed near the expected distribution median, i.e., $p_+ \not\approx p_-$, we found that the distribution of $\hat{\alpha}$ loses properties of the Normal distribution in \eqref{eq:AR_distribution}. Empirical results for observed $\hat{\alpha}$ and $\mathrm{Var}[\hat{\alpha}]$ considering the case when $p_+ \not\approx p_-$ can be found in Appendix \ref{sec:Appendix}.
\end{remark}

By leveraging the distribution of the estimated alarm rate in \eqref{eq:AR_distribution}, bounds of the alarm rate can be made.

\begin{lemma}
    Assuming an uncompromised system with a CUSIGN detector \eqref{pro:CUSIGN_pos} with a reference point $z^{\textrm{ref}}$ and threshold $\tau \in \N^+$, detection of sensor attacks occurs when $\tau^{\alpha}_{-} \leq \hat{\alpha} \leq \tau^{\alpha}_{+}$ where
    \begin{equation}
    \label{eq:lemma2}
    \tau^{\alpha}_{\pm} = \mathrm{E}[\alpha] \pm Z \sqrt{\frac{\theta \mathrm{E}[\alpha](1-\mathrm{E}[\alpha])}{\ell}}.
    \end{equation}

\end{lemma}

\begin{pf}
    Given a CUSIGN detector with threshold $\tau \in \N^+$ and reference point $z^{\textrm{ref}} \in \R^{>0}$ that determine transition probabilities $p_-$ and $p_+$, an expected alarm rate $\mathrm{E}[\alpha]$ can be computed by inverting the first element in \eqref{eq:mu}. With $\mathrm{E}[\alpha]$ and leveraging the Memoryless Run-time Estimator with a pseudo-window of length $\ell$, the distribution of the estimated alarm rate follows $\hat{\alpha} \sim \mathcal{N}(\cdot , \cdot)$ with properties from \eqref{eq:Expected_alpha}. Detection bounds $\tau^{\alpha}_{\pm}$ of a specific confidence level determined by $Z$ of a Normally distributed random variable with properties from \eqref{eq:Expected_alpha} follow $\mathrm{E}[\alpha]~-~Z\sqrt{\frac{\theta \mathrm{E}[\alpha](1~\hspace*{-2pt}-\hspace*{-2pt}~\mathrm{E}[\alpha])}{\ell}}\hspace*{-2pt}~\leq\hspace*{-2pt}~\hat{\alpha}\hspace*{-2pt}~\leq~\hspace*{-2pt} \mathrm{E}[\alpha]~\hspace*{-2pt}+~\hspace*{-2pt}~Z\sqrt{\frac{\theta \mathrm{E}[\alpha](1-\mathrm{E}[\alpha])}{\ell} }$
    satisfying \eqref{eq:lemma2}, concluding the proof.
\end{pf}

Detection of sensor attacks occur when an estimated alarm rate $\hat{\alpha}$ goes beyond a threshold from $\tau^{\alpha}_{\pm}=\{ \tau^{\alpha}_-,\tau^{\alpha}_+ \}$. Lower bounds resulting in $\tau^{\alpha}_- < 0$ are omitted as $\hat{\alpha} \in [0,\frac{1}{\tau}]$.

\section{CUSUM Detector Review}
\label{sec:CUSUM}

The CUSIGN detector alone may not be sufficient as an attacker can change the magnitude of a measurement, but still maintain random signed behavior of the test measure $z_k$. The non-parametric quality of CUSIGN results in the inability to monitor the magnitude of the test measure. A well-known dynamic detector, the \textit{CUmulative SUM} (CUSUM) detector, leverages the magnitude of the test measure sequence $z_k$ to look for changes in the mean from an expectation. Formalized into a model-based attack detector by [\cite{CUSUM_Journal}] that outputs an alarm, the CUSUM attack detection procedure follows 
\vspace{5pt}

\centerline{\textbf{CUSUM Detector Procedure}}
\vspace{3pt}
\hrule
\vspace{-2.5pt}
  \begin{equation}
  \label{pro:CUSUM}
      \begin{array}{ll}
        \hspace*{-7pt} \textbf{Initialize } C_{0} = 0, & \\
        \hspace*{-7pt} C_{k} = \max(0,C_{k-1}+ z_{k} -  b), & \textbf{ if } C_{k-1} \leq \tau^C, \\
        \hspace*{-7pt} C_{k} = 0 \text{ and Alarm } \zeta_{k}^C = 1, & \textbf{ if } C_{k-1} > \tau^C.
      \end{array}
  \end{equation}
  \vspace{-1pt}
  \hrule

\vspace{1pt}
The working principle of this detector is to accumulate the test measure \eqref{eq:z_scalar} in $C_{k}$, triggering an alarm $\zeta_{k}^C = 1$ when the test variable surpasses the threshold $\tau^C$. The test variable $C_{k}$ resets to zero either when the threshold $\tau^C$ is surpassed or when $C_{k}$ goes negative. A bias $b$ is selected based on properties of \eqref{eq:Residual_Covariance} such that $C_{k}$ does not grow unbounded. A detailed explanation of how to construct a transition matrix for the probability distribution $z_k - b$ for the model-based CUSUM can be found in [\cite{CUSUM_Journal}]. The authors provide a method for tuning the threshold $\tau^C$ given a bias $b$ for a desired alarm rate $\mathrm{E}[\alpha^{C}]$ with an assumption that the system is free of sensor attacks, where the residual follows $\bm{r}_k \sim \mathcal{N}(0,\bm{\Sigma})$, hence a shifted $\chi^2$ distribution $z_k-b = \bm{r}_k^T \bm{\Sigma}^{-1} \bm{r}_k - b$.

Considering CUSUM as a stand-alone detector, an adversarial wants to avoid attacks such that the test variable $C_k$ exceeds threshold $\tau^C$ at a higher rate, thereby causing a reset $C_k \hspace{-1pt} = \hspace{-1pt} 0$ in \eqref{pro:CUSUM} by satisfying the CUSUM procedure sequence $C_{k} \hspace{-1pt} = \hspace{-1pt} \max(0,C_{k-1} \hspace{-1pt} + \hspace{-1pt} z_k \hspace{-1pt} - \hspace{-1pt} b) \hspace{-1pt} \leq \hspace{-1pt} \tau^C$ to trigger alarms more often, resulting in a higher alarm rate $\alpha^C$. Subsequently, an attacker can design an attack such that it remains within bounds of CUSUM to not trigger alarms more than expected. To include this attack vector, we can rewrite the CUSUM procedure such that 
\vspace{1pt}
\begin{equation}
    \label{eq:CUSUM_attack_seq}
    C_{k} = \max(0,C_{k-1}+ \big( \| \bm{\Sigma}^{-\frac{1}{2}}(\bm{Ce}_k + \bm{\eta}_k + \bm{\xi}_k) \|^2 \big) -  b).
    \vspace{-2pt}
\end{equation}

Assuming that a malicious attacker can have access to the sensor measurements $\bm{y}_k = \bm{Cx}_k + \bm{\eta}_k$ and has perfect knowledge of the state estimator, it will be able to find the estimator output $\bm{C}\hat{\bm{x}}_k$. With this information, the attacker can solve for $\bm{y}_k - \bm{C}\hat{\bm{x}}_k = \bm{Ce}_k + \bm{\eta}_k$ to achieve the ability of manipulating elements of $\bm{\xi}_k$ by
\begin{equation}
    \label{eq:CUSUM_attack_vector}
    \bm{\xi}_k = - \bm{Ce}_k - \bm{\eta}_k + \bm{\Sigma}^{ \frac{1}{2}} \bm{\xi}^{\tau^C}_k
\end{equation}

such that $\max(0,C_{k-1} + (\bm{\xi}^{\tau^C}_k)^T \bm{\xi}^{\tau^C}_k -b) \leq \tau^C$ can maintain the test variable $C_k$ within the detection threshold $\tau^C$.

\begin{section}{Results} \label{sec:Results}

The proposed CUSIGN detector was validated in simulation and augmented with CUSUM introduced in Section~\ref{sec:CUSUM}. The case study presented in this paper is an autonomous way-point navigation of a skid-steering differential-drive UGV with the following linearized model [\cite{vehiclemodel}]:
\vspace{-1.5pt}
\begin{equation}
\begin{split}
\label{eq:UGV_dynamics}
    \dot{v} &= \frac{1}{m}(F_l+F_r-B_rv), \\
    \dot{\omega} &= \frac{1}{I_z}\Big(\frac{w}{2}(F_l-F_r)-B_l\omega \Big), \text{ } \dot{\theta}_h = \omega,
\end{split}
\end{equation}

where $v$, $\theta_h$, and $\omega$ denotes the velocity, heading angle, and angular velocity, forming the state vector $\bm{x} = [v,\theta_h,\omega]^T$. $F_l$ and $F_r$ describe the left and right input forces from the wheels, $w$ is the vehicle width, while $B_r$ and $B_l$ are resistances due to the wheels rolling and turning. The continuous-time model \eqref{eq:UGV_dynamics} is discretized with a sampling rate $t_s = 0.01$ to satisfy the system model described in \eqref{eq:system1}. The UGV is tasked to continuously navigate to four goal-points along a square trajectory with side lengths of $5$m maintaining a velocity $v=0.5$m/s for $200$s.

In the simulation, we perform two different attack sequences on the velocity sensor on-board the vehicle: 1) a persistent attack and 2) an alternating pattern attack. Both stealthy attack sequences are designed to be undetectable by CUSUM, but are detected by CUSIGN due to the creation of non-random patterns.

\subsection{Simulations}
\label{sec:simulation}

We first consider the system under normal conditions where $\bm{\xi}_k = 0$. In Table \ref{table:AR} we show the alarm rate of the system over $5$ million data samples and compare the results to the expected alarm rate $\mathrm{E}[\alpha^{\pm}]$ computed from \eqref{eq:mu} in the case where $p_+ = p_- = 0.5$ for thresholds $\tau = 1,2,3,4$. Next, in Fig.~\ref{fig:empirical_alpha_hat} we show the distribution of the alarm rate estimate $\hat{\alpha}$ from the four cases in Table~\ref{table:AR} overlaid with the expected distributed curve (in red) according to \eqref{eq:Expected_alpha}.

\begin{table}[htb!]
\caption{$\mathrm{E}[\alpha ^{\pm}]$ when $p_+ = p_- = 0.5$.}
\vspace{-6pt}
\centering
\begin{tabular}{ |p{1.5cm}||p{1.1cm}|p{1.1cm}|p{1.1cm}|p{1.1cm}|  }
 \hline
 \rule{0pt}{1.2\normalbaselineskip} Thresholds  & \centering $\tau = 1$ & \centering $\tau = 2$ & \centering $\tau = 3$ & \hspace*{3pt} $\tau = 4$ \\[3pt]
 \hline
 \rule{0pt}{.9\normalbaselineskip} \centering $\mathrm{E}[\alpha^{\pm}] $   & \centering $0.5$ & \centering $0.1 \bar{6}$ & \centering $0.08 \bar{3}$ & \hspace*{7pt} $0.05$ \\[.2pt]
 \hline
 \rule{0pt}{.9\normalbaselineskip} \centering $\alpha^{\pm}$ (sim.)   & \centering $.50006$ & \centering $.16692$ & \centering $.083291$ & \hspace*{-2pt} $.050012$ \\[.4pt]
 \hline
\end{tabular}
\label{table:AR}
\end{table}

\begin{figure}[htb!]
\subfigure[\label{fig:tau1} ]{\includegraphics[width = 0.115\textwidth]{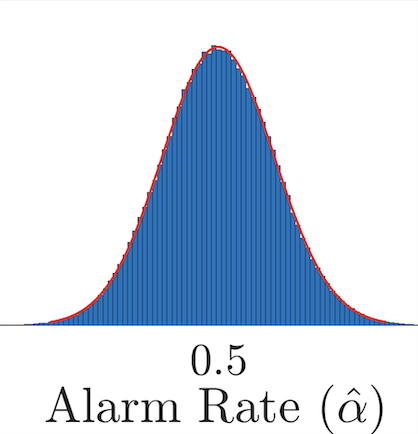}}
\subfigure[\label{fig:tau2} ]{\includegraphics[width = 0.115\textwidth]{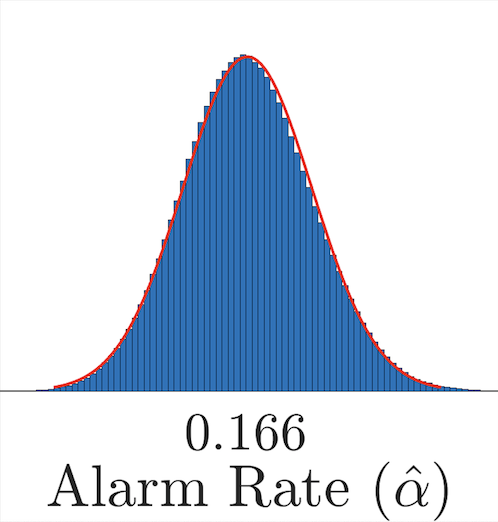}}
\subfigure[\label{fig:tau3} ]{\includegraphics[width = 0.115\textwidth]{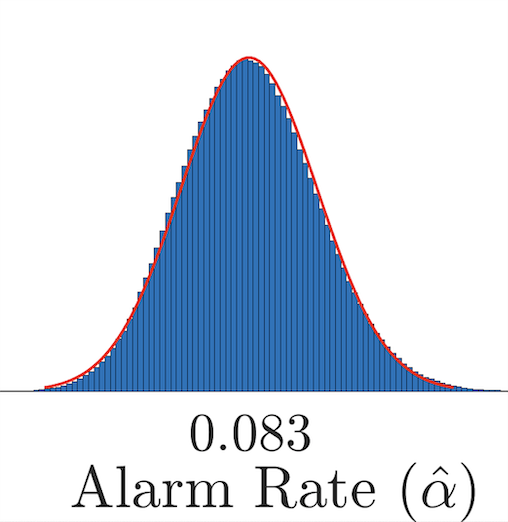}}
\subfigure[\label{fig:tau4} ]{\includegraphics[width = 0.115\textwidth]{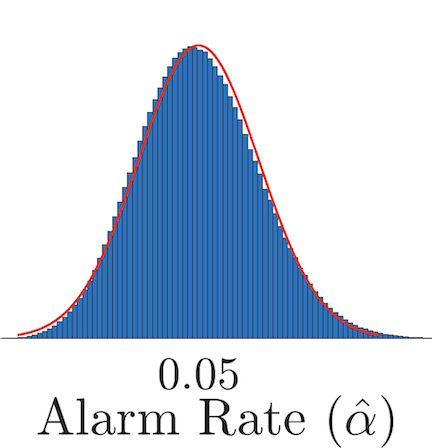}}
\vspace{-8pt}
\caption{Resulting distributions of $\hat{\alpha}$ when $p_+ = p_- = 0.5$ for (a) $\tau = 1$, (b) $\tau = 2$, (c) $\tau = 3$, (d) $\tau = 4$.}
\label{fig:empirical_alpha_hat}
\end{figure}

Now, considering the UGV \eqref{eq:UGV_dynamics} case study in the presence of hidden attacks on the velocity sensor on state $x_1 = v$, we show the detection capabilities of CUSIGN. The CUSIGN is designed with $z^{\textrm{ref}} = \mathrm{E}[\text{median}(z_k)] \approx s \big(1-\frac{2}{9s} \big)^3$ where $s=3$ such that the transition probabilities satisfy $p_{\pm}=0.5$ and threshold $\tau = 2$. The expected alarm rate $\mathrm{E}[\alpha] = 0.1 \bar{6}$ and the Memoryless Run-time Estimator \eqref{eq:alpha_est1} with pseudo-window length $\ell = 100$ has detection bounds \eqref{eq:lemma2} at $\tau^{\alpha}_{-} = 0.0987$ and $\tau^{\alpha}_{+} = 0.2347$ where $Z=3$ for a $ 99.7 \%$ confidence. The design of CUSUM contains a bias $b = 1.1s = 3.3$ with a threshold $\tau^C = 2.3226$ to satisfy an expected alarm rate $\mathrm{E}[\alpha^C] = 0.15$ (see [\cite{CUSUM_Journal}] for tuning details), where the alarm rate is computed by a conventional method of length $\ell$ by $\frac{1}{\ell} \sum^k_{k-\ell+1} \zeta^C_k$. Fig. \ref{fig:attack1_results} shows the results of a persistent attack \eqref{eq:CUSUM_attack_seq}, \eqref{eq:CUSUM_attack_vector} beginning at $k=10,000$ with a noiseless magnitude of $0.1\tau^C$. The alarm rate $\hat{\alpha}^C$ for CUSUM is unaffected while CUSIGN discovers the attack and alarm rates $\hat{\alpha}^{\pm}$ both go beyond the detection bounds $\tau^{\alpha}_{\pm}$ (red dashed lines). A second attack shown in Fig. \ref{fig:attack2_results} is attempted with an alternating noiseless pattern of $\{ 0.1\tau^C,-0.1\tau^C \}$ to show that CUSIGN can detect patterns. Again, alarm rates for CUSIGN find the non-random patterns and go beyond the detection bounds $\tau^{\alpha}_{\pm}$ while CUSUM is not able to detect the non-random behavior.

\begin{figure}[b!ht]
\centering
\includegraphics[width=.99\linewidth]{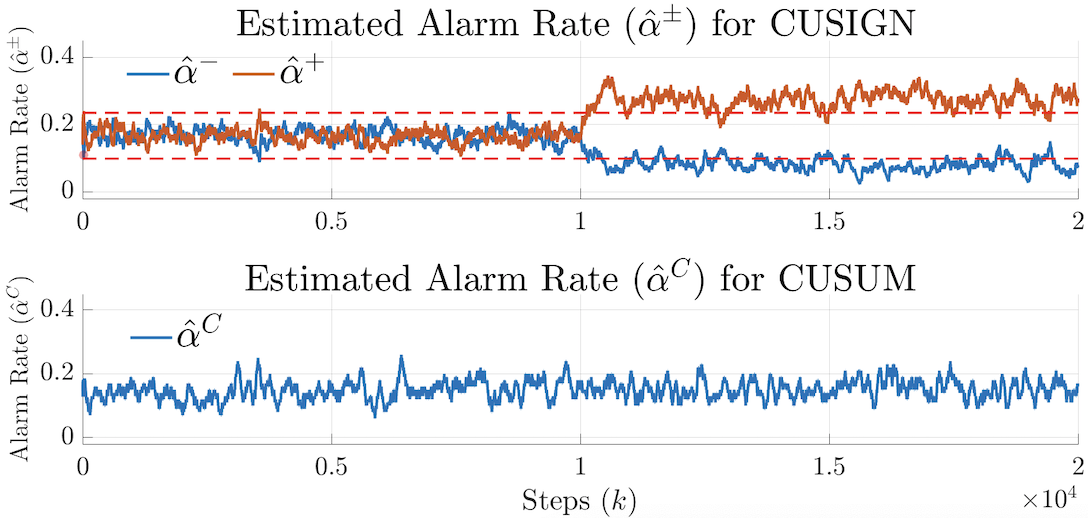}
\vspace{-13pt}
\caption{Alarm rates $\hat{\alpha}^{\pm}$ and $\hat{\alpha}^{C}$ for both CUSIGN and CUSUM with a hidden persistent sensor attack.}
\label{fig:attack1_results}
\end{figure}

\begin{figure}[b!ht]
\centering
\includegraphics[width=.99\linewidth]{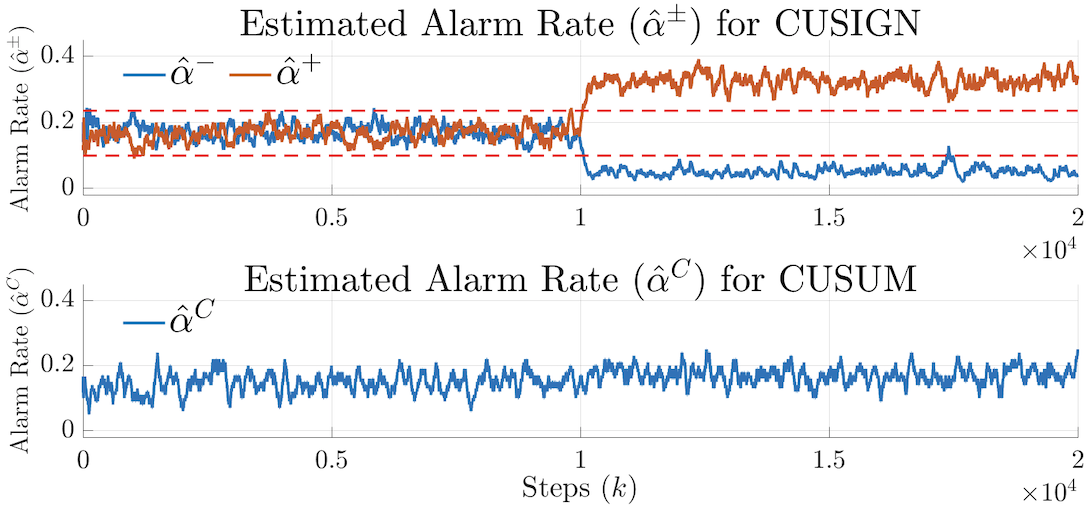}
\vspace{-13pt}
\caption{Alarm rates $\hat{\alpha}^{\pm}$ and $\hat{\alpha}^{C}$ for both CUSIGN and CUSUM with a hidden alternating sensor attack.}
\label{fig:attack2_results}
\end{figure}
\vspace{7pt}

\end{section}
\begin{section}{Conclusions \& Future Work} 
\label{sec:conclusion}
\vspace{-1pt}

In this paper we have characterized the CUSIGN procedure for detection of hidden sensor attacks that present non-random behavior. In particular, we have constructed a Markov chain of the CUSIGN test sequence to model a resulting expected alarm rate. We have formalized a memoryless run-time method for computing an alarm rate estimate using a modified version of Welford's online algorithm with a pseudo-window, which we call the Memoryless Run-time Estimator (MRE). We empirically found the resulting estimated alarm rate distribution and leveraged it to provide detection bounds given a specific level of confidence. Then, we characterized attack sequences that remain undetected to the CUSUM dynamic attack detector, that leave trails of non-random behavior for CUSIGN to detect the attack.

In our future work we plan to extend the current work to leverage CUSIGN on CPSs with redundant sensors to detect and remove compromised sensors and create an attack resilient controller.

\end{section}

\bibliography{ms}             % bib file to produce the bibliography

\appendix
\begin{section}{Empirical Results} \label{sec:Appendix}
\vspace{-1pt}

From Remark \ref{rem:remark1} in Section \ref{sec:AR_memoryless}, we show in Fig. \ref{fig:empirical_alpha} the gradual divergence from the normal approximation as $p_+$ and $p_-$ are no longer similar as the distribution of the estimated alarm rate estimate $\hat{\alpha}$ becomes skewed. The empirical results provided throughout this section are results from $5$ million samples, thus giving an accurate representation of the resulting distributions.

\begin{figure}[htb!]
\subfigure[\label{fig:a} ]{\includegraphics[width = 0.16\textwidth]{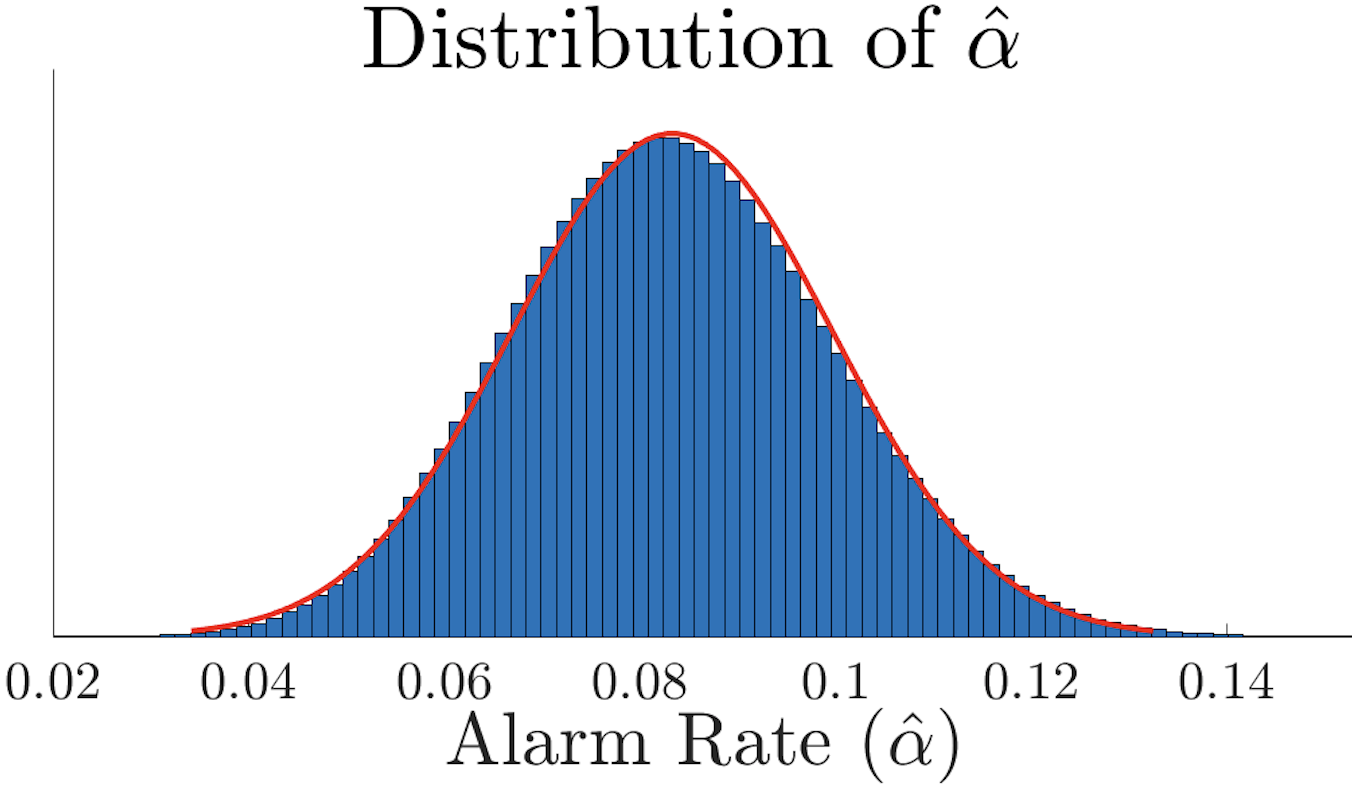}}
\subfigure[\label{fig:b} ]{\includegraphics[width = 0.16\textwidth]{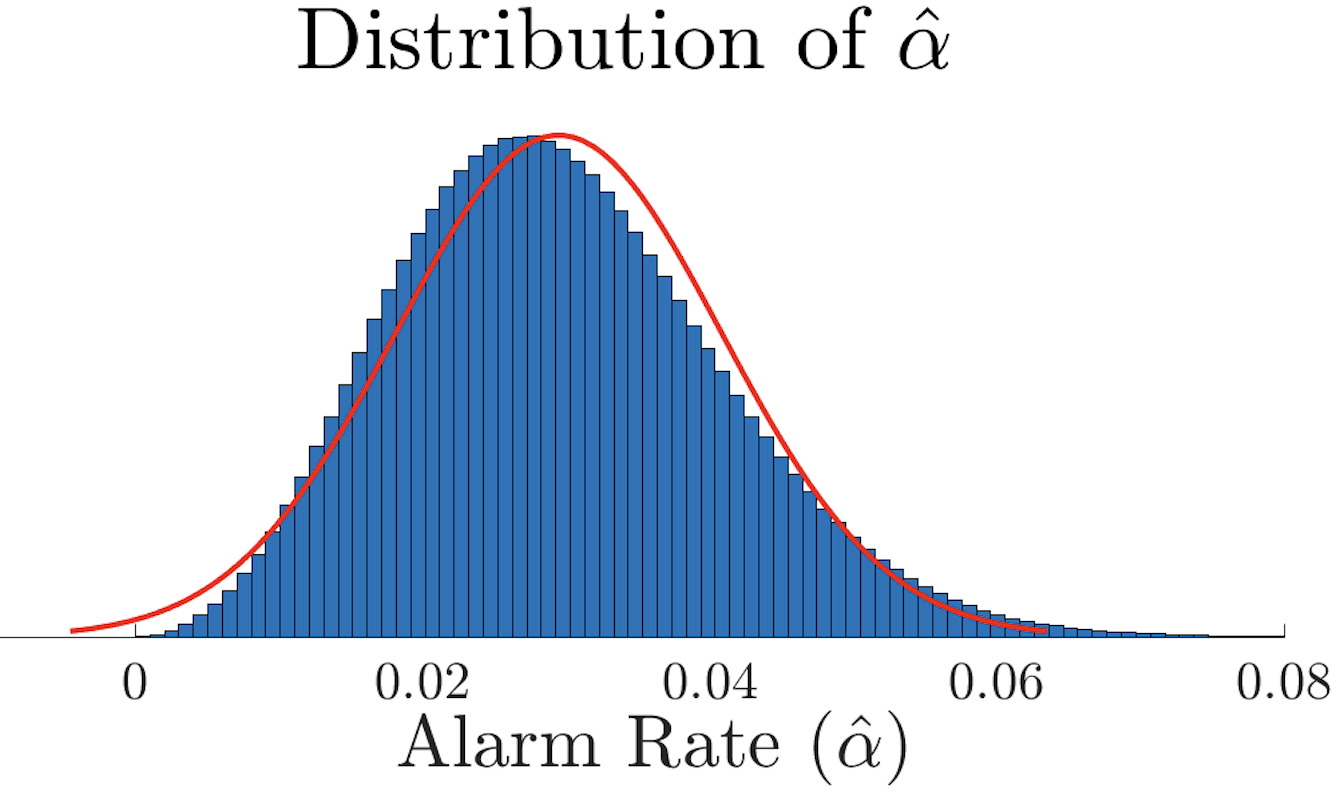}}
\subfigure[\label{fig:c} ]{\includegraphics[width = 0.153\textwidth]{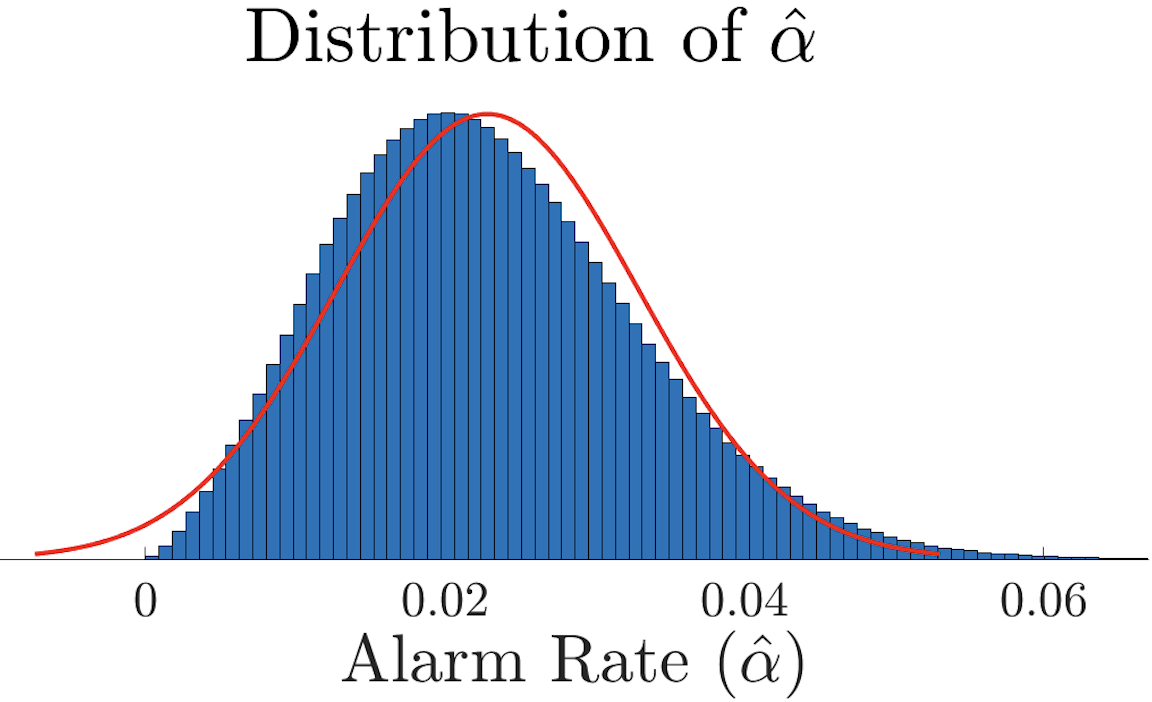}}
\vspace{-10pt}
\caption{Resulting distributions of $\hat{\alpha}$ when (a) $p_+ = p_- = 0.5$, (b) $p_- = 0.37$, (c) $p_- = 0.3$.}
\label{fig:empirical_alpha}
\end{figure}

Furthermore, Table \ref{table:mean_alpha_hat2} provides the expected $\mathrm{E}[\hat{\alpha}]$ and simulated alarm rates, while Table \ref{table:var_alpha_hat2} provides the square root of the expected and simulated variance $\sqrt{\mathrm{Var}[\hat{\alpha}]}$ (i.e., standard deviation). It can be seen that as $p_{\pm} \approx 0.5$, the simulated mean of the alarm rate estimates remain approximately equal to the expectation (i.e., $ \bar{\hat{\alpha}} \approx \mathrm{E}[\alpha]$), but the simulation results for standard deviation diverge from the expected variance as $p_+ \not\approx p_-$.

\begin{table}[htb!]
\caption{Results of $\mathrm{E}[\hat{\alpha}]$ for $\ell = 100$.}
\vspace{-7pt}
\centering
\begin{tabular}{ |p{2.35cm}||p{1.55cm}|p{1.55cm}|p{1.55cm}|  }
 \hline
 \rule{0pt}{1.2\normalbaselineskip} \centering $p_{\pm} $  & \centering $.4$ & \centering $.5$ & \hspace*{15pt} $.6$ \\[2pt]
 \hline
 \rule{0pt}{.9\normalbaselineskip} \centering $ \mathrm{E}[\hat{\alpha}]/\text{sim} \text{ } (\tau \hspace*{-2pt} = \hspace*{-2pt} 1)$   & \centering $.400/4.01$ & \centering $.500/.500$ & \hspace*{1.25pt} $.600/6.01$ \\[.8pt]
 \hline
 \rowcolor{VeryLightGray}
 \rule{0pt}{.9\normalbaselineskip} \centering $\mathrm{E}[\hat{\alpha}]/\text{sim} \text{ } (\tau \hspace*{-2pt} = \hspace*{-2pt} 2)$   & \centering $.1143/.1142$ & \centering $.166\bar{6}/.1665$ & \hspace*{-2.7pt} $.2250/.2251$ \\[.8pt]
 \hline
 \rule{0pt}{.9\normalbaselineskip} \centering $\mathrm{E}[\hat{\alpha}]/\text{sim} \text{ } (\tau \hspace*{-2pt} = \hspace*{-2pt} 3)$   & \centering $.0484/.0483$ & \centering $.083\bar{3}/.0832$ & $.1256/.1258$ \\[.8pt]
 \hline
 \rowcolor{VeryLightGray}
 \rule{0pt}{.9\normalbaselineskip} \centering $\mathrm{E}[\hat{\alpha}]/\text{sim} \text{ } (\tau \hspace*{-2pt} = \hspace*{-2pt} 4)$   & \centering $.0244/.0239$ & \centering $.0500/.0500$ & $.0835/.0833$ \\[.8pt]
 \hline
\end{tabular}
\label{table:mean_alpha_hat2}
\end{table}

\begin{table}[htb!]
\caption{Results of $\mathrm{std}[\hat{\alpha}]$ for $\ell = 100$.}
\vspace{-7pt}
\centering
\begin{tabular}{ |p{2.35cm}||p{1.55cm}|p{1.55cm}|p{1.55cm}|  }
 \hline
 \rule{0pt}{1.2\normalbaselineskip} \centering $p_{\pm} $  & \centering $.4$ & \centering $.5$ & \hspace*{15pt} $.6$ \\[2pt]
 \hline
 \rule{0pt}{.9\normalbaselineskip} \centering $ \mathrm{std}[\hat{\alpha}]/\text{sim} \text{ } (\tau \hspace*{-2pt} = \hspace*{-2pt} 1)$   & \centering $.0346/.0347$ & \centering $.0354/.0355$ & \hspace*{-2.55pt} $.0346/.0347$ \\[.8pt]
 \hline
 \rowcolor{VeryLightGray}
 \rule{0pt}{.9\normalbaselineskip} \centering $\mathrm{std}[\hat{\alpha}]/\text{sim} \text{ } (\tau \hspace*{-2pt} = \hspace*{-2pt} 2)$   & \centering $.0194/.0204$ & \centering $.0227/.0226$ & \hspace*{-2.7pt} $.0254/.0238$ \\[.8pt]
 \hline
 \rule{0pt}{.9\normalbaselineskip} \centering $\mathrm{std}[\hat{\alpha}]/\text{sim} \text{ } (\tau \hspace*{-2pt} = \hspace*{-2pt} 3)$   & \centering $.0127/.0138$ & \centering $.0163/.0163$ & $.0196/.0185$ \\[.8pt]
 \hline
 \rowcolor{VeryLightGray}
 \rule{0pt}{.9\normalbaselineskip} \centering $\mathrm{std}[\hat{\alpha}]/\text{sim} \text{ } (\tau \hspace*{-2pt} = \hspace*{-2pt} 4)$   & \centering $.0091/.0099$ & \centering $.0128/.0128$ & $.0163/.0153$ \\[.8pt]
 \hline
\end{tabular}
\label{table:var_alpha_hat2}
\end{table}

\end{section}

\end{document}